# Pitchfork Bifurcation and Zharov Splitting in Nonlinear Mid-infrared Photothermal Spectroscopy in a liquid crystal using a Quantum Cascade Laser

Alket Mertiri[1], Hatice Altug[5,6,7], M. K. Hong[3,6], P. Mehta[3], J. Mertz[4,6] , L. D. Ziegler[2,6] and Shyamsunder Erramilli[3,4,6]

[1]Division of Materials Science and Engineering, Boston University, Boston, Massachusetts, 02215, USA

[2]Department of Chemistry, 590 Commonwealth Ave, Boston, Massachusetts, 02215, USA

[3]Department of Physics, Boston University, Boston, Massachusetts, 02215, USA

[4]Department of Biomedical Engineering, Boston University, Boston, Massachusetts, 02215, USA

[5]Department of Electrical and Computer Engineering, Boston, Massachusetts, 02215, USA

[6]Photonics Center, 8 St Mary's St, Boston, Massachusetts, 02215, USA

[7]Department of BioEngineering, Ecole Polytechnique Federale De Lausanne, Lausanne, CH-1015, SWITZERLAND

Correspondence: shyam@bu.edu

## Abstract

We report on the mid-infrared nonlinear photothermal spectrum of the neat liquid crystal 4-Octyl-4'-Cyanobiphenyl (8CB) using a tunable Quantum Cascade Laser (QCL). The nonequilibrium steady state characterized by the nonlinear photothermal infrared response undergoes a supercritical pitchfork bifurcation. The bifurcation, observed in heterodyne two-color pump-probe detection, leads to ultrasharp nonlinear infrared spectra similar to those reported in the visible region. A systematic study of the peak splitting as function of absorbed infrared power shows the bifurcation has a critical exponent of 0.5. The surprising observation of an apparently universal critical exponent in a nonequilibrium state is explained using a simple model reminiscent of mean field theory. Apart from the intrinsic interest for nonequilibrium studies, nonlinear photothermal methods lead to a dramatic narrowing of spectral lines, giving rise to a potential new contrast mechanism for the rapidly emerging new field of mid-infrared microspectroscopy using QCLs.

Photothermal spectroscopy[1] has rapidly emerged as the most sensitive label-free optical spectroscopic method rivaling fluorescence spectroscopy particularly for nonradiative excited states. A nonequilibrium state is created in a sample by the absorption of a modulated pump laser, and detected using the scatter[2] of a probe laser. When the pump laser is pulsed periodically, phase-lock methods allow the resultant nonequilibrium steady state to be studied with high sensitivity. In linear photothermal spectroscopy, the scattered probe signal is a linear function of the pump power, with reported sensitivity down to the single molecule level at room temperature[3,4]. Recognition of the advantages of sensitivity and label-free nature have led to rapid development of linear photothermal methods in the visible region, both for spectroscopy and imaging nanoparticles and organelles[5-9]. Photothermal spectroscopy has been used to characterize weak absorption[9-12] in solids and liquids, and to measure chemical kinetics in solution[13,14], and for spectroscopy of heme proteins and imaging in mitochondria[13,15-18]. Zharov[19-22] and co-workers have extended photothermal and photoacoustic spectroscopy to the nonlinear regime, and have reported splitting and sharpening of photothermal spectral signatures in the visible. Nonlinear spectral methods have the potential to significantly enhance the range of applications of photothermal microscopy[23], and superresolution imaging[24]. We extend nonlinear photothermal methods into the mid-infrared region of the spectrum using a Quantum Cascade Laser as the pump laser, and present a detailed study of the dramatic peak splitting phenomenon, called *Zharov splitting*, characterized by a sharpening of mid-infrared photothermal spectrum in a liquid crystal. The tunability and power control of the QCL allow us to study the peak splitting in greater detail than was possible before. We show that the peak splitting is analogous to a *supercritical bifurcation* in nonlinear dynamical systems, with a "mean field" exponent of 0.5. The observations raise the surprising possibility of universal behavior in the nonequilibrium steady state.

Mid-infrared photothermal spectroscopy is attractive because of characteristic vibrational molecular normal modes, especially in the so-called "fingerprint" region[23]. With the invention of tunable Quantum Cascade Lasers (QCL), mid-infrared spectroscopy[25-28] is poised for rapid growth. The spectral brightness of these table-top QCL sources exceeds that of synchrotrons and other large relativistic electron-accelerator-based sources[29]. Indium Antimonide (InSb) or Mercury-Cadmium-Telluride (MCT)[19] liquid-nitrogen cooled detectors are intrinsically less sensitive than the best available photodetectors in the visible. A further attraction of photothermal spectroscopy is that it facilitates the use of sensitive room-temperature optical detectors for infrared spectroscopy. Recently, we have reported on QCL-based photothermal heterodyne mid-infrared spectroscopy[30,31]. Farahi et al reported on a homodyne QCL-based method for remote sensing[32].

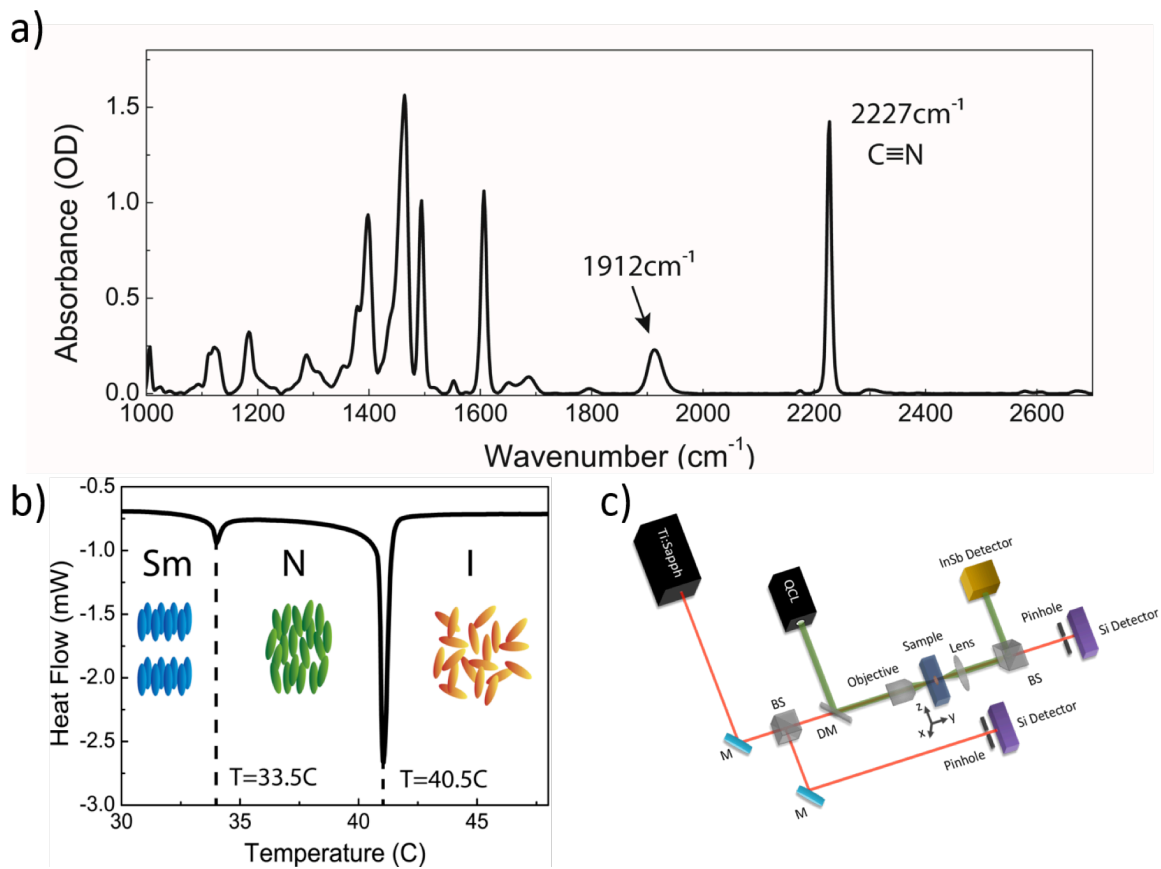

FIG. 1. (a) FTIR absorption spectrum of 8CB. (b) Differential scanning calorimetry of 8CB liquid crystal showing three phases with an illustration of the orientations of the molecule director in the Smectic-A phase (Sm); the Nematic phase (N); and the Isotropic phase(I). (c) Experimental setup for photothermal detection. The modulated QCL pump beam and Ti:Sapphire probe beam are co-aligned with the dichroic mirror and focused onto the sample by a special ZnSe objective. M-mirror, DM-dichroic mirror, and BS-beamsplitter.

The experimental setup for both nonlinear and linear photothermal IR spectroscopy is shown in Fig. 1(c). A tunable Quantum Cascade Laser source tuned to a selected intrinsic absorption band serves as the modulated pump beam[30]. Absorption in the sample causes a thermally induced nonequilibrium change $\Delta n$ in the refractive index experienced by a collinear probe beam, tuned to a wavelength far from any resonance. The probe beam, provided by a Ti:Sapphire laser tuned to 800 nm in cw mode, experiences scattering due to $\Delta n$, which can be detected by measuring the modulated scattered probe intensity[33], either in homodyne detection with the resultant signal being proportional to $\Delta n^2$, or in heterodyne detection where the signal is proportional to $\Delta n$.

Cyanobiphenyls form a well-studied class of thermotropic liquid crystals, with rich phase behavior[34,35,36]. The liquid crystal 4-Octyl-4'-Cyanobiphenyl (8CB) experiences well-known phase transitions, from smectic-A phase (SmA) to nematic phase (N) at 306.5 K and then to the isotropic phase (I) at 313.5 K[37-39] shown by the endotherms in differential scanning calorimetry (DSC) measurements in Fig. 1(b). The FTIR absorbance spectrum[36] in Fig. 1(a) shows a weak combination band, thought to arise from out-of-plane CH vibrations[40] centered at 1912 cm[-1] (also Fig S1) and lies within the tuning range of the QCL laser[30] with a molar extinction coefficient of 14.9 M[-1] cm[-1]. The base temperature of the sample is controlled using a circulating water bath. The photothermal signal due to the weak mid-infrared combination mode (Fig 1(a)) could be observed in all of the smectic, nematic and isotropic phases of 8CB. Previous optical photothermal studies of liquid crystals required the use of a dye[41] or gold nanoparticles[42], and a detailed understanding of guest-host interactions[43]. In contrast, photothermal infrared spectroscopy[30-32] is inherently label-free.

**Results**

Fig 2 shows a sequence of 25 photothermal absorption spectra of 8CB at a temperature of 29 C in the smectic-A phase as the current in the QCL, and hence the incident IR power, is increased (see

Methods). The intensity at the sample varied over a range 0.1-1.2 ×10⁴ W/cm². The observed photothermal signal depends on the power absorbed by the sample at each frequency. For a given value of the QCL current, the incident power at each frequency was independently measured using the cryogenically cooled detector, and calibrated using a power meter in the absence of the sample cell. The protocol for acquiring the data is described in *Methods*. At low pump power, the photothermal spectrum is identical to the linear IR absorption spectrum (Fig S2(a)). The linear spectrum was measured in two different ways – using a commercial FTIR spectrometer and by measuring the transmitted IR signal from the QCL using an InSb cryogenically cooled detector in single beam mode. The IR signal was measured simultaneously along with the photothermal response. Also shown in Fig 2 is the calculation from a phenomenological mean-field model for the nonlinear photothermal response described in detail in "Results". Figure 3 shows the transition in the photothermal response between the linear and nonlinear regime in greater detail.

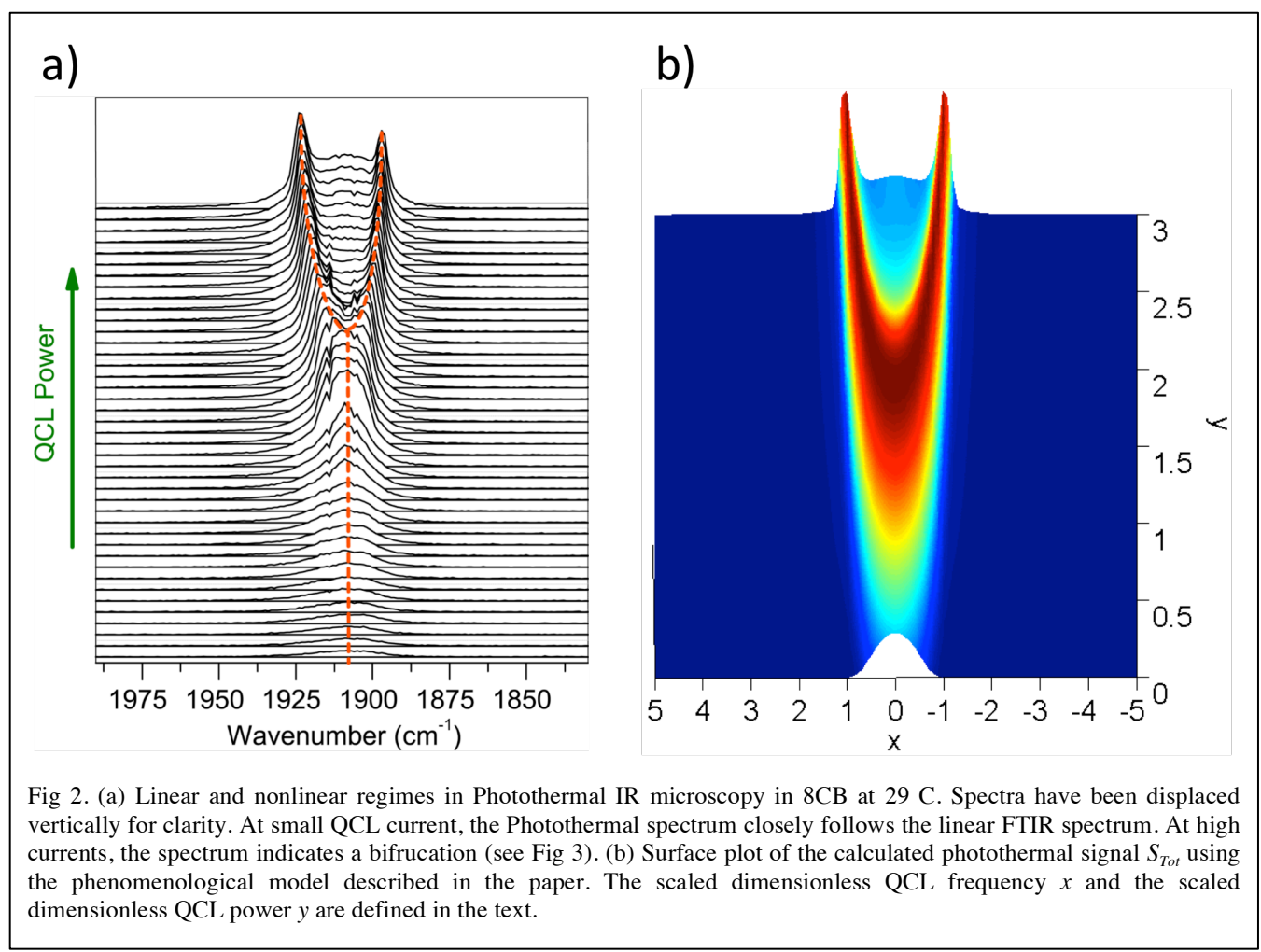

Fig 2. (a) Linear and nonlinear regimes in Photothermal IR microscopy in 8CB at 29 C. Spectra have been displaced vertically for clarity. At small QCL current, the Photothermal spectrum closely follows the linear FTIR spectrum. At high currents, the spectrum indicates a bifurcation (see Fig 3). (b) Surface plot of the calculated photothermal signal $S_{Tot}$ using the phenomenological model described in the paper. The scaled dimensionless QCL frequency $x$ and the scaled dimensionless QCL power $y$ are defined in the text.

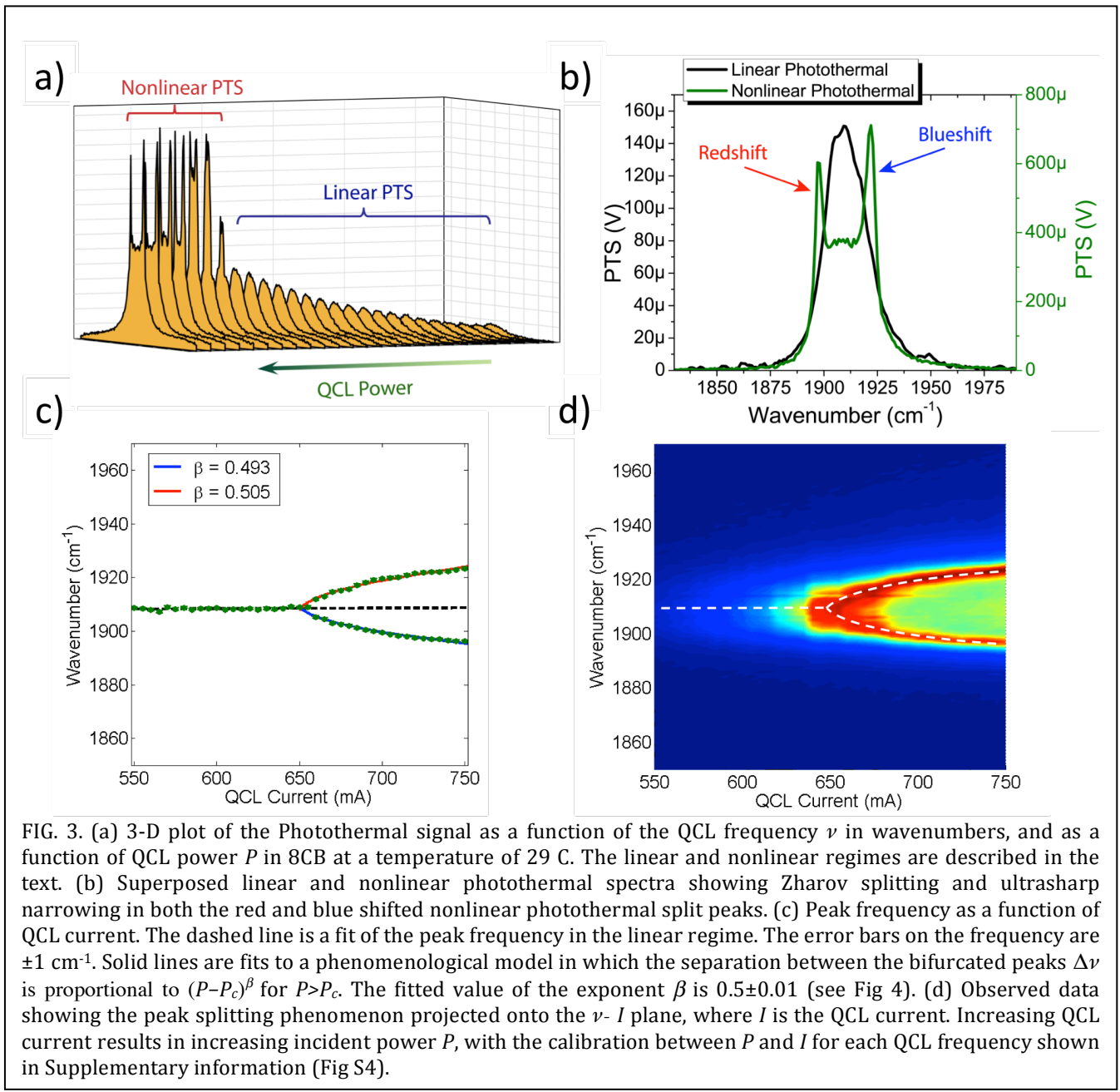

FIG. 3. (a) 3-D plot of the Photothermal signal as a function of the QCL frequency $\nu$ in wavenumbers, and as a function of QCL power $P$ in 8CB at a temperature of 29 C. The linear and nonlinear regimes are described in the text. (b) Superposed linear and nonlinear photothermal spectra showing Zharov splitting and ultrasharp narrowing in both the red and blue shifted nonlinear photothermal split peaks. (c) Peak frequency as a function of QCL current. The dashed line is a fit of the peak frequency in the linear regime. The error bars on the frequency are $\pm 1$ cm$^{-1}$. Solid lines are fits to a phenomenological model in which the separation between the bifurcated peaks $\Delta \nu$ is proportional to $(P-P_c)^\beta$ for $P>P_c$. The fitted value of the exponent $\beta$ is $0.5\pm0.01$ (see Fig 4). (d) Observed data showing the peak splitting phenomenon projected onto the $\nu$- $I$ plane, where $I$ is the QCL current. Increasing QCL current results in increasing incident power $P$, with the calibration between $P$ and $I$ for each QCL frequency shown in Supplementary information (Fig S4).

Four different regimes can be identified: (i) As the QCL current is increased, the magnitude of the PTS signal at the peak absorption frequency 1912 cm$^{-1}$ increases linearly at first, but the band shape is unchanged. (ii) At higher currents, it can be seen that the amplitude of the PTS also has a weak quadratic dependence (Fig 3(a)), but the shape of the spectral band is still unaltered. (iii) Above a first threshold evident in Fig 3, a narrow spike appears centered near the absorption peak frequency. (iv) As the absorbed power is increased beyond a critical threshold of absorbed power $P_c$, the narrow peak splits into two branches, and each branch sharpens still further. Fig 3(b) shows a superposition of the photothermal spectrum in the linear and nonlinear regimes. The spectra are normalized to illustrate the dramatic narrowing of the split red and blue shifted peaks in the nonlinear spectrum, compared to the linear absorption spectrum. The separation $\Delta\nu$ between the peak center frequencies of each of the two branches increases with power $P$. We find that $\Delta\nu \propto (P-P_c)^{1/2}$ (Fig 3 (c)-(d)). For a given base sample temperature $T$, the critical threshold absorbed power $P_c$ required for peak splitting is remarkably reproducible, and is a function $T$.

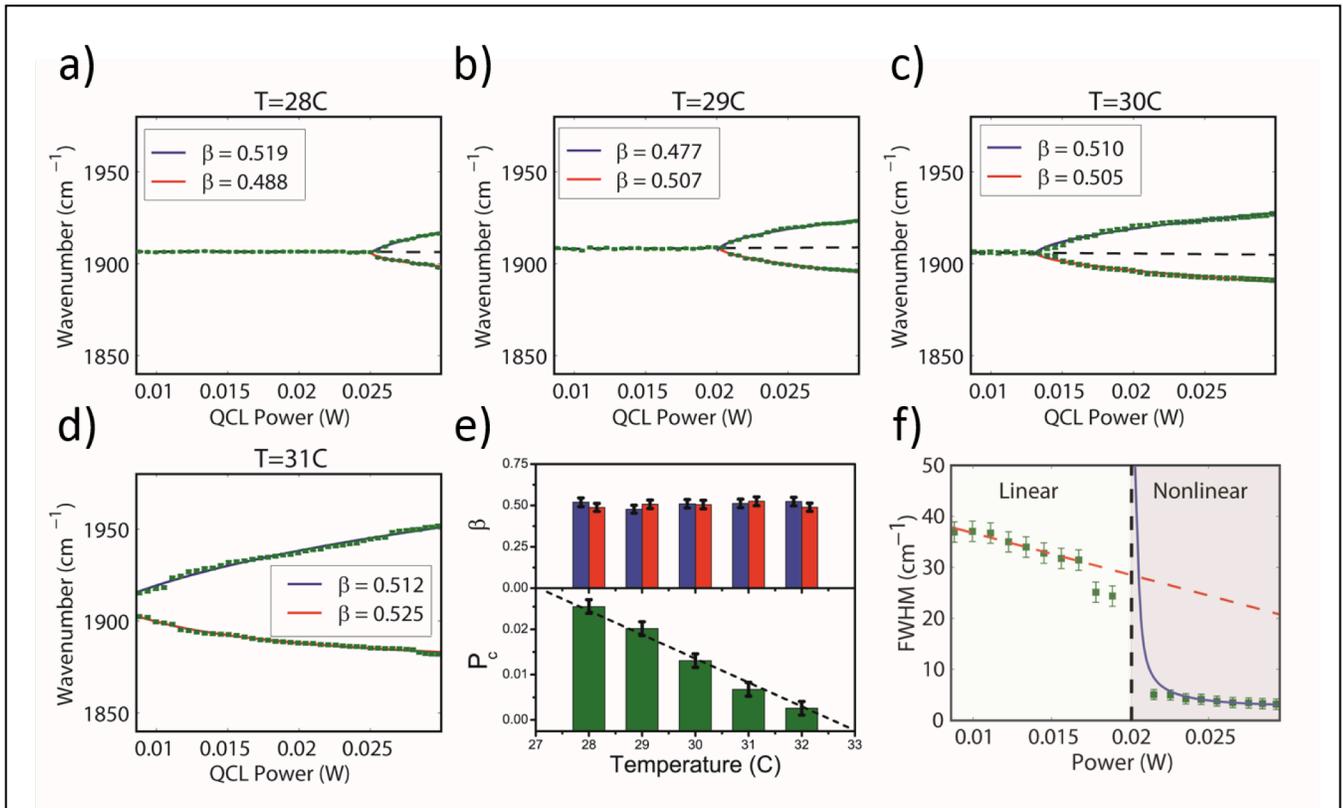

FIG. 4. (a-d) Analysis of nonlinear Photothermal bifurcation at different sample temperatures 28 C – 31 C. Data points are obtained as in Fig 3, and the fits are solid lines $\Delta\nu$ is proportional to $(P-P_c)^\beta$ where the exponent $\beta$ was allowed to vary. (e) Histogram of the experimentally derived exponents at different temperatures for both the blue-shifted(blue bars) and red-shifted (red bars) branches of the bifurcation. The exponent $\beta = 0.5\pm0.01$. Also shown is a histogram of the fitted values of $P_c$ at different temperatures. (f) Full-Width-at-Half-Maximum (FWHM) of the linear PTS spectrum, and the blue-shifted nonlinear branch following bifurcation, at 29 C. The solid line is a fit to the phenomenological model with the FWHM is proportional to $(P-P_c)^{-1/2}$.

**Temperature dependence.** In order to investigate the nature of the bifurcation, we measured the photothermal response as a function of temperature, QCL frequency and absorbed power. Figure 4 shows the peak splitting response for a selected set of sample base temperatures (Fig S3). At each temperature, the peaks split, and display the same $\Delta\nu \propto (P-P_c)^{1/2}$ dependence on the incident QCL laser power $P$. The robustness of the scaling exponent was tested by fitting to a generalized power law of the form $\Delta \nu = b \left( P - P_c \right)^{\beta}$ where the exponent $\beta$ is allowed to vary and is not constrained (Fig 4 (a-e)), along with the parameter $b$. A histogram of the exponents under different experimental conditions gave a range for $\beta = 0.5 \pm 0.01$, independent of temperature (Fig 4 d). The critical power decreased linearly as the sample temperature is raised from 28 C to 32 C, approaching the SmA-N phase transition temperature (Fig 4e). Above about 31 C, peak splitting occurs in the photothermal spectrum even just above the lasing threshold of the QCL. To summarize, all the peak splitting data can be fitted with a single universal curve in which the peak frequency split can be fit to form $\Delta \nu = b \left( P - P_c \right)^{\beta}$, with an apparently universal temperature independent exponent $\beta = 0.5$.

**Discussion**

While it is initially tempting to conclude that our observations listed above are specific to phase transitions in liquid crystals, the observation of the peak splitting phenomenon by Zharov and co-workers on a completely different aqueous physical system suggests that the phenomenon is rather general, requiring only the nucleation of a higher temperature phase. The exponent associated with the peak splitting phenomenon is observed in our work for a *nonequilibrium* system, complementary to equilibrium studies in the rich literature on effective critical exponents belonging to the 3-D XY universality class near the smectic A-nematic phase transition in bulk 8CB[44], and in other systems[35,45]. The observed exponent of 0.5 is generally expected for dynamical systems that exhibit pitchfork bifurcation[46], which is intimately related to the mean field dynamics theory[47,48] of an *equilibrium* order parameter with $Z_2$ symmetry, such as the Ising model near a phase transition. The framework for analysis of our nonequilibrium steady state experiments is set by the prior work on nanoparticles in aqueous media in the visible region of the electromagnetic spectrum[19], where the peak splitting effect was attributed to reduced forward scattering due to the formation of "nanobubbles" caused by laser heating. The formation of bubbles, with a discontinuous boundary in the refractive index between two phases, results in significant elastic backscatter of the probe beam due to a combination of Mie scattering and multiple scattering events. The contribution to forward elastic scatter of the probe beam then decreases, and the observed PTS signal also decreases. Optical scattering from thermally generated clusters[19] thus reduces the photothermal response in the nonlinear regime. A simple way to think about this rather complex explanation is the following: the PTS signal is due to the formation of a thermal lens[1]; creation of bubbles destroys the thermal lens leading to a loss in the signal[19]. Such a mechanism for peak splitting is expected to hold for mid-infrared photothermal spectroscopy as well. Based on this analysis, we expect that the first observable reduction in the scattered photothermal signal to coincide with the peak of the infrared pump absorption, i.e. at 1912 cm⁻¹, as shown in Fig 3(b). The microscopic mechanism for bubble or cluster generation, and the scattering cross-section, varies from sample to sample. The observed peak splitting phenomenon shares some general aspects of spectral hole burning[49], but the approach here is qualitatively different.

In classical hole burning the imaginary part of the dielectric function at the probe frequency is altered, (i.e. reduced absorption). In our case the real part of the dielectric response at the probe frequency, related to the refractive index, is altered. In support of this hypothesis, we note below that the spectrum measured by the cryogenic infrared detector does *not* show peak splitting (Fig S5); only the scattered photothermal response does.

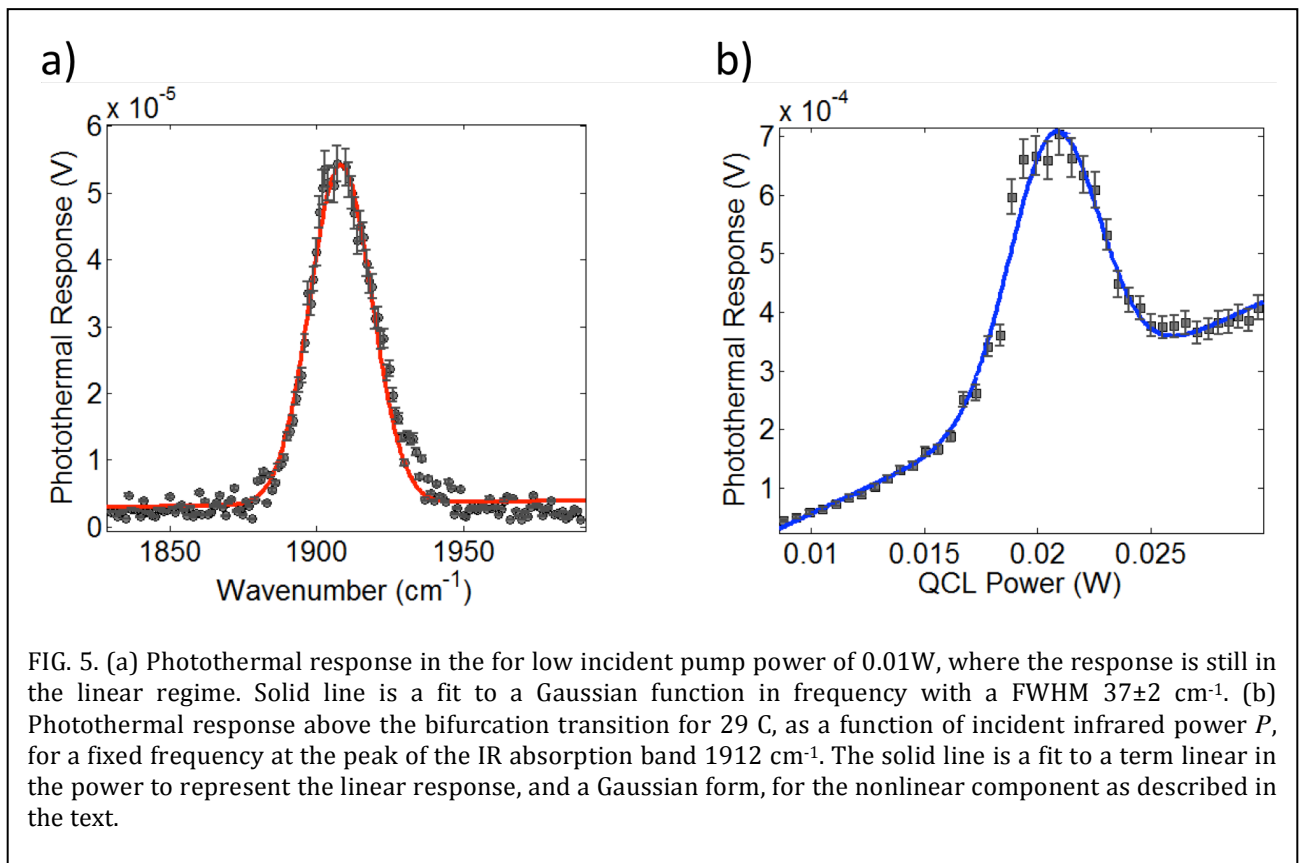

FIG. 5. (a) Photothermal response in the for low incident pump power of 0.01W, where the response is still in the linear regime. Solid line is a fit to a Gaussian function in frequency with a FWHM 37±2 cm⁻¹. (b) Photothermal response above the bifurcation transition for 29 C, as a function of incident infrared power $P$, for a fixed frequency at the peak of the IR absorption band 1912 cm⁻¹. The solid line is a fit to a term linear in the power to represent the linear response, and a Gaussian form, for the nonlinear component as described in the text.

To understand why peak splitting occurs in the photothermal spectrum, but not in the IR absorption spectrum that is being simultaneously monitored, the discussion in the above paragraph suggests a simple model. The nonlinear contribution $S_{NL}$ to the observed photothermal scattering signal in Fig 3(a) can be considered to be a function of two variables $S_{NL}(\nu, P)$, where $\nu$ is the infrared QCL frequency and $P$ is the incident power of the QCL that serves as control parameter. The dependence on other experimental variables like the temperature, probe wavelength, probe power is implicit in the parametrization of the model. At a fixed QCL frequency close to the peak of the infrared band, the observed nonlinear contribution can be fit to a phenomenological Gaussian functional form, as shown in Fig 5(a).

$$S_{NL}(\nu,P) = S_0 e^{-\xi^2[P-P_T(\nu)]^2} \quad \text{Eq (1)}$$

For a fixed QCL frequency, the parameter $\xi$ sets the width of the Gaussian as a function of incident QCL power. The parameter $P_T$ depends on the IR QCL frequency, and has a simple interpretation: it represents the incident QCL power at which the maximum photothermal signal is detected. Physically it represents a threshold power above which scatter from nanobubble formation leads to a decrease in the forward scattered probe signal. Since the bubble formation depends on the power *absorbed* in the sample, we expect $P_T$ to be the smallest when the QCL is tuned to the peak of the resonance frequency ~ 1912 cm$^{-1}$. As the QCL frequency is de-tuned from resonance, increasingly greater incident power is needed to reach the same threshold for bubble formation. It is then obvious that the functional dependence of $P_T$ on the frequency is the reciprocal of the *linear* FTIR absorbance spectrum. For the specific case of the combination band studied here, the linear IR spectrum is approximately a Gaussian centered at the peak IR frequency ~ $e^{-(\nu-\nu_0)^2/2\sigma^2}$, where $\sigma$ is determined by the Full-Width-at-Half-Maximum (FWHM) of the linear FTIR spectrum. The threshold parameter in Eq(1) is then given by $P_T(\nu) = P_c e^{+(\nu-\nu_0)^2/2\sigma^2}$ leading to a simple analytic functional form for the nonlinear photothermal signal:

$$S_{NL}(\nu,P) = S_0 \exp\left[-\xi^2\left\{P - P_c e^{+(\nu-\nu_0)^2/2\sigma^2}\right\}^2\right] \quad \text{Eq (2)}$$

In the simplest analysis, the total experimentally observed signal $S_{Tot}$ is a sum of the nonlinear component $S_{NL}$, and the linear response $S_L$, $S_{Tot} = S_L + S_{NL}$. In this analysis the linear response is of the form $S_L(\nu,P) = f_\lambda P e^{-(\nu-\nu_0)^2/2\sigma^2}$ where the factor $f_\lambda$ depends on the probe wavelength, probe power, collection and geometric factors, as discussed extensively by Berciaud et al[2]. In simply summing the linear and nonlinear responses, interference effects are neglected. Fig 5b shows a surface plot of the sum of the linear and nonlinear contributions $S_{Tot} = S_L + S_{NL}$. As seen in Fig 5b, the function captures the essence of the peak splitting phenomenon, and yields the correct value of the critical exponent 0.5. $S_0$ is independent of the QCL frequency. The peak frequency of the nonlinear photothermal spectrum $S_{NL}$ can be obtained easily from the analytic form. To make contact with well-established formalism of nonlinear maps and dynamical systems[46], we consider the formal problem of finding the maximum of function $S_{NL}(\nu, P)$. Defining a "free energy" $F = -\ln S_{NL}(\nu, P)$, the maxima are then the minima of $F$ at constant power $P$, given by $\left. \frac{\partial F}{\partial \nu} \right|_P = 0$. The roots of the resulting algebraic equation give the peak frequency. Peak splitting occurs when the equation has multiple roots. Considering the functional form as a nonlinear dynamical map, expressed in terms of dimensionless variables $x, y$

$$x = \left(\nu - \nu_0\right)^2 / 2\sigma^2, y = P / P_c \Rightarrow S = S_0 \exp\left[-\zeta^2 \left(y - e^{x^2}\right)^2\right] \Rightarrow F = \zeta^2 \left(y - e^{x^2}\right)^2 \qquad \text{Eq(3)}$$

Close to the maximum $x=0$, the map is clearly quadratic

$$F = \zeta^2 \left(y - 1 - x^2 + ...\right)^2 \Rightarrow F / \zeta^2 \approx \left(y - 1\right)^2 - 2\left(y - 1\right)x^2 + x^4$$

$$\left. \frac{\partial F}{\partial x} \right|_y = 0 \Rightarrow -4\left(y - 1\right)x + 4x^3 = 0 \Rightarrow \begin{cases} x = 0 \\ x_\pm = \pm\sqrt{y - 1} \end{cases} \qquad \text{Eq(4)}$$

Expressed in terms of the observed variables $P$ and QCL frequency, it can be seen that (i) when $P$ is less than $P_c$ the peak frequency is at $\nu_0$ which is the maximum of the linear FTIR spectrum; (ii) when $P$ exceeds $P_c$ the peak frequency of the nonlinear scattering signal is split into two branches, with $\Delta\nu = b\left(P - P_c\right)^\beta$ as observed. The exponent 0.5 arises mainly from the symmetry. The Gaussian nature of the functional forms are not important, so long as they belong to the universality class of quadratic maps[46].

*Relationship between $P_c$ and the equilibrium phase transition*. As stated above, at low power, the PTS signal is primarily due the thermal lens effect, i.e. due primarily to the change in the refractive index induced by the Gaussian beam near the focal spot. The threshold power $P_c$ required to create a bubble of the higher temperature phase depends on how far the sample temperature is from the SmA-N phase transition temperature. $P_c$ decreases as the temperature approaches the SmA-N transition temperature. Thus the parameter $P_c$, which is determined in the non-equilibrium steady state, is linked closely to an underlying equilibrium phase transition[48].

*Limitations of the approach*: The "mean field" approach also correctly predicts that the sharpening of the observed nonlinear photothermal spectral peaks above the pitchfork bifurcation. Fig 3 shows that the widths of the nonlinear peaks, estimated by the Full-Width-at-Half-Maximum (FWHM), decrease as the incident QCL power increases. A straightforward calculation of the FWHM using the model of Eq(2) confirms this. Quantitatively, just above the bifurcation transition, each of the two bifurcated peaks has a line width that diverges as $\Delta x_{\pm} = \left\langle \left( x - x_{\pm} \right)^2 \right\rangle^{1/2} \propto \left( y - 1 \right)^{-\beta'}$ where the exponent $\beta'$ is also 0.5. Re-casting this in terms of the experimental unscaled variables, the width of the narrow peaks decreases with increasing QCL power $P$ as $\left( P - P_c \right)^{-1/2}$. The width of the ultrasharp features in our experiments is determined primarily by the $\sim 1$ cm$^{-1}$ linewidth and repeatibility of the QCL laser frequency, and the precision with which the QCL power can be controlled. Studies of linewidths suggest the limitations of the approach. While Fig 3(b) confirms the sharpening experimentally, there is an observed asymmetry in the line-widths in the upper and lower branches of the bifurcation, in studies performed at fixed QCL current. In the data shown in Fig 3, the red-shifted peak has a linewidth of 1.7 cm$^{-1}$, compared to the 2.5 cm$^{-1}$ linewidth of the blue-shifted peak. The asymmetry is due to several reasons. The first is a systematic experimental effect because, even at constant QCL current, the IR power incident at the two peak frequencies $\nu_{\pm}$ are slightly different (Fig S4), because of a small offset between the maximum of the frequency in the Gain curve of the QCL and the 1912 cm$^{-1}$ IR absorption band peak. Correction for this systematic experimental effect does not however completely remove the observed asymmetry. A second contribution is due to an underlying weak asymmetry in the IR absorption spectrum (Fig S1(b)), which is not taken into account in Eq(2). Extension to non-symmetric vibrational spectra may be done by expressing the threshold power in terms of the appropriate spectral moments, and will lead to corrections for the exponent. The analysis here must therefore be regarded as a first step, with extensions necessitated by further observations on other nonlinear photothermal systems, accounting for fluctuations and finite size effects in systems driven into non-equilibrium steady state.

The pitchfork bifurcation and dramatic sharpening of infrared absorption spectroscopy may be universal with applications to biological systems and materials science and engineering.

## Methods

**Vibrational Infrared Photothermal Spectrometer.** In our setup, the pump and probe beams are collinearly combined using a dichroic mirror (DM) and focused coaxially onto the sample by a zinc-selenide (ZnSe) focusing objective (NA=0.25). The mid-IR Quantum Cascade Laser (Daylight Solutions) is focused onto the sample with a Gaussian beam waist diameter of 22 $\mu$m ± 3 $\mu$m. Losses at the beamsplitters and in the ZnSe objective lens resulted in an estimated QCL pump beam incident intensity on the sample of up to a maximum of ~ 1.2 ×10$^4$ W/cm$^2$. The probe beam is separated by a beamsplitter into a reference beam and a sample beam and the photothermal signal is measured as their ratio. At the sample, the probe beam has a beam waist diameter of 16 $\mu$m ± 3 $\mu$m. The incident probe beam power was set at 100 mW. Losses in the ZnSe objective and the coupling optics, due primarily to reflection losses, resulted in an estimated probe beam intensity at the sample of ~ 2×10$^4$ W/cm$^2$. The transmitted pump and probe beams are then collected by a ZnSe lens. The pump and probe beams are separated using a second dichroic beamsplitter. The pump beam is focused onto an InSb liquid nitrogen-cooled detector for in situ monitoring of the IR absorption. The intensity of the probe beam is measured using a Si-photodetector. This set-up allows for comparison between direct mid-infrared detection and heterodyne photothermal detection in the same sample under identical conditions (Fig 1). The QCL was operated in pulse mode, with 500 ns pulses with a repetition rate of up to 100 kHz, corresponding to a maximum duty cycle of 5%. To enhance sensitivity, the probe signal is detected using a conventional Si photodetector using a lock-in amplifier, phase locked to the modulation frequency of the pump beam.

**QCL Infrared spectroscopy.** For mid-IR photothermal studies, the 4-Octyl-4'-Cyanobiphenyl (8CB) liquid crystal sample was sandwiched between cleaned CaF$_2$ windows with 50 $\mu$m Mylar spacer. In the absence of rubbing or surface coating the molecular alignment of the bulk is homogeneous, with no preferred direction at either CaF$_2$ window substrate. Observation with visible light under crossed polarizers did not show homeotropic alignment of the sample as a whole. QCL mid IR laser beam was tuned across the absorption peak of the sample, centered at 1912 cm$^{-1}$. As the QCL current was increased from the threshold of 500 mA to a maximum of 750 mA, the average output power varied over a range from ~ 5 mW to above 30 mW. The base temperature of the sample was controlled by using a circulating water bath and measured with a thermocouple integrated into the brass sample holder.

**Phothermal heterodyne scattering**. In linear photothermal spectroscopy, the signal depends on both the pump power $P_{pump}$ and the probe power $P_{probe}$, the absorption spectrum of the sample as a function of the pump frequency, as well as on geometric factors. In the theoretical models developed by Berciaud et al[2], for a sample with a characteristic absorption spectrum described by $f(v)$, the linear scattering signal power $S_{Linear}$ depends on the probe wavelength $\lambda_{probe}$ and is linearly proportional to both the probe laser power, and the pump power.

In our nonlinear experiments, the observed signals are still linear in the probe power, but nonlinear as a function of the pump power. All the data presented in the paper used a lock-in amplifier to detect the scattered probe signal modulated at the pump frequency of 100 kHz. The transmitted pump power provides an *in situ* measurement of the infrared absorption spectrum. The transmitted IR power is consistent with the FTIR linear spectrum, and does not show a bifurcation transition.

**Line-widths of Nonlinear peaks**. The narrow line-width of each branch above the bifurcation can be estimated using elementary arguments, by calculating the second moment in the neighborhood of each peak. The FWHM $\Gamma_+$ the blue-shifted branch with a peak at $x_+$ decreases with increasing power as $(P-P_c)^{-1/2}$. The FWHM $\Gamma_-$ of the red-shifted branch is equal to that of the blue-shifted branch in the simplest model.

Given $x_+ = +\sqrt{1-y}$ the width $\Gamma_+ = \left\langle \left(x-x_+\right)^2 \right\rangle = \dfrac{\displaystyle\int_{x_+-\Delta}^{x_+ +\Delta} \left(x-x_+\right)^2 S_0 \exp\left[-\zeta^2\left(y-e^{x^2}\right)^2\right]dx}{\displaystyle\int_{x_+-\Delta}^{x_+ +\Delta} S_0 \exp\left[-\zeta^2\left(y-e^{x^2}\right)^2\right]dx}$

For $\Delta \ll \left|x_+ - x_-\right|$ and for $y>1$ we have $\left\langle \left(x-x_+\right)^2 \right\rangle = \dfrac{1}{4\zeta^2\left(y-1\right)} + O\left(\Delta^2\right) \Rightarrow$

Full-width-Half-Maximum $\Gamma_+ \propto \left(P-P_c\right)^{-1/2}$ and similarly for $\Gamma_- = \Gamma_+$

## Acknowledgments

We acknowledge support from NIH grant number 1 R21 EB013381-01, and NSF I/UCRC grant number NSF IIP-1068070. We thank T. Jeys and V. Liberman, MIT Lincoln Labs, for discussions and the generous loan of the QCL system.

# References

1    Bialkowski, S. E. *Photothermal Spectroscopy Methods for Chemical Analysis*. 584 (1996).

2    Berciaud, S., Lasne, D., Blab, G. A., Cognet, L. & Lounis, B. Photothermal heterodyne imaging of individual metallic nanoparticles: Theory versus experiment. *Phys Rev B* **73**, 045424 (2006).

3    Armani, A. M., Kulkarni, R. P., Fraser, S. E., Flagan, R. C. & Vahala, K. J. Label-free, single-molecule detection with optical microcavities. *Science* **317**, 783-787, (2007).

4    Gaiduk, A., Yorulmaz, M., Ruijgrok, P. V. & Orrit, M. Room-Temperature Detection of a Single Molecule's Absorption by Photothermal Contrast. *Science* **330**, 353-356, doi:10.1126/science.1195475 (2010).

5    Lasne, D. *et al.* Label-free optical imaging of mitochondria in live cells. *Opt Express* **15**, 14184-14193, doi:144192 [pii] (2007).

6    Lu, S. Label-free imaging of heme proteins with two-photon excited photothermal lens microscopy. *Appl. Phys. Lett.* **96**, 113701 (2010).

7    Cognet, L. *et al.* Single metallic nanoparticle imaging for protein detection in cells. *Proceedings of the National Academy of Sciences* **100**, 11350-11355, doi:10.1073/pnas.1534635100 (2003).

8    Boyer, D., Tamarat, P., Maali, A., Lounis, B. & Orrit, M. Photothermal Imaging of Nanometer-Sized Metal Particles Among Scatterers. *Science* **297**, 1160-1163, doi:10.1126/science.1073765 (2002).

9    Lima, S. M. *et al.* Mode-mismatched thermal lens spectrometry for thermo-optical properties measurement in optical glasses: a review. *J Non-Cryst Solids* **273**, 215-227, doi:Doi 10.1016/S0022-3093(00)00169-1 (2000).

10   Cabrera, H. *et al.* Measurement of the Soret coefficients in organic/water mixtures by thermal lens spectrometry. *Cr Mecanique* **341**, 372-377, doi:Doi 10.1016/J.Crme.2013.01.011 (2013).

11   Gupte, S. S., Marcano, O., Pradhan, R. D., Desai, C. F. & Melikechi, N. Pump-probe thermal lens near-infrared spectroscopy and Z-scan study of zinc (tris) thiourea sulfate. *J Appl Phys* **89**, 4939-4943, doi:Doi 10.1063/1.1358325 (2001).

12   Bhattacharyya, I., Kumar, P. & Goswami, D. Probing Intermolecular Interaction through Thermal-Lens Spectroscopy. *J Phys Chem B* **115**, 262-268, doi:Doi 10.1021/Jp1062429 (2011).

13   Proskurnin, M. A., Nedosekin, D. A. & Kuznetsova, V. V. Investigation of Belousov-Zhabotinsky reaction kinetics using thermal lens spectrometry. *Rev Sci Instrum* **74**, 343-345, doi:Doi 10.1063/1.1519674 (2003).

14   Astrath, N. G. C. *et al.* Thermal-lens study of photochemical reaction kinetics. *Opt Lett* **34**, 3460-3462 (2009).

15   Brusnichkin, A. V. *et al.* Ultrasensitive label-free photothermal imaging, spectral identification, and quantification of cytochrome c in mitochondria, live cells, and solutions. *J Biophotonics* **3**, 791-806, doi:Doi 10.1002/Jbio.201000012 (2010).

16   Nedosekin, D. A., Galanzha, E. I., Ayyadevara, S., Reis, R. J. S. & Zharov, V. P. Photothermal confocal spectromicroscopy of multiple cellular chromophores and fluorophores (vol 102, pg 672, yr 2012). *Biophys J* **102**, 1235-1235, doi:Doi 10.1016/J.Bpj.2012.02.012 (2012).

17      Nedosekin, D. A., Galanzha, E. I., Ayyadevara, S., Reis, R. J. S. & Zharov, V. P. Photothermal Confocal Spectromicroscopy of Multiple Cellular Chromophores and Fluorophores. *Biophys J* **102**, 672-681, doi:Doi 10.1016/J.Bpj.2011.12.035 (2012).

18      Tong, L. & Cheng, J. X. Label-free imaging through nonlinear optical signals. *Mater Today* **14**, 264-273 (2011).

19      Zharov, V. P. Ultrasharp nonlinear photothermal and photoacoustic resonances and holes beyond the spectral limit. *Nat Photonics* **5**, 110-116 (2011).

20      Zharov, V. P. & Lapotko, D. O. Photothermal imaging of nanoparticles and cells. *Ieee J Sel Top Quant* **11**, 733-751, doi:Doi 10.1109/Jstqe.2005.857382 (2005).

21      Zharov, V. P., Galitovsky, V. & Viegas, M. Photothermal detection of local thermal effects during selective nanophotothermolysis. *Appl Phys Lett* **83**, 4897-4899, doi:Doi 10.1063/1.1632546 (2003).

22      Khodakovskaya, M. V. *et al.* Complex genetic, photothermal, and photoacoustic analysis of nanoparticle-plant interactions. *P Natl Acad Sci USA* **108**, 1028-1033, doi:Doi 10.1073/Pnas.1008856108 (2011).

23      Rajakarunanayake, Y. N. & Wickramasinghe, H. K. Nonlinear Photothermal Imaging. *Appl Phys Lett* **48**, 218-220, doi:Doi 10.1063/1.96800 (1986).

24      Nedosekin, D. A., Galanzha, E. I., Dervishi, E., Biris, A. S. & Zharov, V. P. Super-Resolution Nonlinear Phototherma Microscopy. *Small*, doi:10.1002/smll.201300024 (2013).

25      Faist, J. *et al.* Quantum Cascade Laser. *Science* **264**, 553-556, doi:Doi 10.1126/Science.264.5158.553 (1994).

26      Capasso, F., Faist, J., Sirtori, C. & Cho, A. Y. Infrared (4-11 um) quantum cascade lasers. *Solid State Communications* **102**, 231-236 (1997).

27      Capasso, F. *et al.* High performance quantum cascade lasers for the lambda=4 to 17 mu m region and their chemical sensing applications. *Conf P Indium Phosph*, 262-265, doi:Doi 10.1109/Iciprm.2000.850282 (2000).

28      Capasso, F. *et al.* New frontiers in quantum cascade lasers and applications. *Ieee J Sel Top Quant* **6**, 931-947 (2000).

29      Carr, G. L., Dumas, P., Hirschmugl, C. J. & Williams, G. P. Infrared synchrotron radiation programs at the National Synchrotron Light Source. *Nuovo Cimento Della Societa Italiana Di Fisica D-Condensed Matter Atomic Molecular and Chemical Physics Fluids Plasmas Biophysics* **20**, 375-395 (1998).

30      Mertiri, A. *et al.* Mid-infrared photothermal heterodyne spectroscopy in a liquid crystal using a quantum cascade laser. *Appl Phys Lett* **101** (2012).

31      Mertiri, A., Hong, M. K., Mertz, J., Altug, H. & Erramilli, S. Mid-Infrared Photothermal Response in a Liquid Crystal using a Quantum Cascade laser. *APS March Meeting Bulletin Y1.00011*, Y1.00011 (2012).

32      Farahi, R. H., Passian, A., Tetard, L. & Thundat, T. Pump-probe photothermal spectroscopy using quantum cascade lasers. *J Phys D: Applied Phyiscs* **45**, 125101 (2012).

33      Berne, B. J. & Pecora, R. *Dynamic Light Scattering*.  (Wiley, 1976).

34      Kumar, S. & Kang, S.-W. in *Encyclopedia of Condensed Matter* Vol. 3  (eds G. Bassani, G. Liedl, & P. Wyder) 111-120 (Elsevier Ltd, Oxford, UK, 2005).

35      Kumar, S. *Liquid Crystals in the Nineties and Beyond*.  (World Scientific, 1995).

36      Thomas, M. Two-dimensional FT-IR correlation analysis of the phase transitions in a liquid crystal, 4′-n-octyl-4-cyanobiphenyl (8CB). *Vibrational Spectroscopy* **24**, 137-146 (2000).

37    Davidov, D. *et al.* High-resolution x-ray and light-scattering study of critical behavior associated with the nematic—smectic-A transition in 4-cyano-4′-octylbiphenyl. *Phys Rev B* **19**, 1657 (1979).

38    Kutnjak, Z., Kralj, S., Lahajnar, G. & Zumer, S. Thermal study of octylcyanobiphenyl liquid crystal confined to controlled-pore glass. *Fluid Phase Equilibria* **222-223**, 275-281, doi:10.1016/j.fluid.2004.06.005 (2004).

39    Thoen, J., Marynissen, H. & Van Dael, W. Temperature dependence of the enthalpy and the heat capacity of the liquid-crystal octylcyanobiphenyl (8CB). *Physical Review A* **26**, 2886 (1982).

40    Frunza, L. *et al.* Surface layer in composites containing 4-n-octyl-4 '-cyanobiphenyl. FTIR spectroscopic characterization. *J Mol Struct* **651**, 341-347 (2003).

41    Yelleswarapu, C. S. *et al.* Phase contrast imaging using photothermally induced phase transitions in liquid crystals. *Appl Phys Lett* **89** (2006).

42    Parra-Vasquez, A. N. G., Oudjedi, L., Cognet, L. & Lounis, B. Nanoscale Thermotropic Phase Transitions Enhance Photothermal Microscopy Signals. *The Journal of Physical Chemistry* **3**, 1400-1403 (2012).

43    Truong, T. V., Xu, L. & Shen, Y. R. Early dynamics of guest-host interaction in dye-doped liquid crystalline materials. *Phys Rev Lett* **90** (2003).

44    Garland, C. W. & Nounesis, G. Criical behavior at nematic-smectic-A phase transitions. *Physical Review E* **49**, 2964-2971 (1994).

45    Garland, C. W. *et al.* Critical behavior at the nematic-smectic-A transition in butyloxybenzylidine heptylaniline (4O.7). *Physical Review A* **27**, 3234-3240 (1983).

46    Strogatz, S. H. *Nonlinear Dynamics and Chaos.*  (Perseus Books, 1994).

47    Ten Bosch, A., Maissa, P. & Sixou, P. A Landau-de Gennes theory of nematic polymers. *Journal de Physique Letters* **44**, 105-111 (1983).

48    Stanley, H. E. *Introduction to Phase Transitions and Critical Phenomena.*  (Oxford University Press, 1971).

49    Mukamel, S. *Principles of Nonlinear Spectroscopy.*  321ff (Oxford University Press, 1995).